\documentclass[10pt]{emulateapj}

\slugcomment{ApJ, in press}

\shorttitle{Knots in the Helix Nebula found in H$_2$}
\shortauthors{Matsuura et al.}

\begin{document}
\title{A `firework' of H$_2$ knots in the Planetary Nebula NGC\,7293 (the Helix Nebula)
\altaffilmark{*}
}

\author{M. Matsuura\altaffilmark{1,2}, 
A.K. Speck\altaffilmark{3},
B.M. McHunu\altaffilmark{3},
I. Tanaka\altaffilmark{4},
N.J. Wright\altaffilmark{2},
M.D. Smith\altaffilmark{5},
A.A. Zijlstra\altaffilmark{6},
S. Viti\altaffilmark{2},
R. Wesson\altaffilmark{2}
}

\altaffiltext{*}{Based on data taken with the Subaru Telescope, National Astronomical Observatory of Japan (proposal ID S07B-054)}
\altaffiltext{1}{National Astronomical Observatory of Japan, Osawa 2-21-1, 
        Mitaka, Tokyo 181-8588, Japan;
       e-mail: m.matsuura@nao.ac.jp}
\altaffiltext{2}{Department of Physics and Astronomy, University College London, 
        Gower Street, London WC1E 6BT, UK;
        e-mail (m.m.): mikako@star.ucl.ac.uk}
\altaffiltext{3}{Physics and Astronomy, University of Missouri, Columbia, MO 65211, USA}
\altaffiltext{4}{Subaru Telescope, National Astronomical Observatory of Japan, 
       650 North Aohoku Place, Hilo, HI 96720, USA}
\altaffiltext{5}{Centre for Astrophysics and Planetary Science,
        University of Kent, Canterbury CT2 7NH, UK}
\altaffiltext{6}{Jodrell Bank Centre for Astrophysics, University of Manchester, Oxford Street,
       Manchester M13 9PL, UK}

\begin{abstract}
We present a deep and wide field-of-view ($4' \times 7'$) image of the
planetary nebula (PN) NGC\,7293 (the Helix Nebula) in the 2.12 $\mu$m H$_2$ $v=1
\rightarrow 0$ S(1) line.  The  excellent seeing (0.4$''$) at
the Subaru Telescope, allows the details of cometary knots to be examined.
The knots are found at distances of 2.2$'$--6.4$'$ from the
central star (CS).  At the inner edge and in the inner ring (up to 4.5$'$ from
the CS), the knot often show a `tadpole' shape, an elliptical head with
a bright crescent inside and a long tail opposite to the CS.  In detail,
there are variations in the tadpole shapes, such as narrowing tails, widening
tails, meandering tails, or multi-peaks within a tail.  In the outer ring
(4.5$'$--6.4$'$ from the CS), the shapes are more fractured, and the
tails do not collimate into a single direction.
The transition in knot morphology from the inner edge to the outer ring is
clearly seen.
The number density of knots governs the H$_2$ surface brightness
in the inner ring: H$_2$ exists only within the knots.
Possible mechanisms which contribute to the shaping of the knots are 
discussed, including photo-ionization and streaming motions. A plausible
interpretation of our images is that inner knots are being overrun by a faster
wind, but that this has not (yet) reached the outer knots.
Based on H$_2$ formation and destruction rates, H$_2$ gas can survive in knots 
from formation during the late asymptotic giant branch (AGB) phase throughout the PN phase. 
These observations provide new constraints on the formation and evolution of
knots, and on the physics of molecular gas embedded within ionized gas.
\end{abstract}




\section{Introduction}

Heterogeneities and filamentary structures in the ionized emission from planetary nebulae (PNe)
have been known for many years \citep[e.g.][]{Curtis1918, Zanstra55,
  Vorontsov-Velyaminov68}.  Such structures are now known to be common in all
types of nebulae, including PNe \citep[e.g.][]{Goncalves2001, Matsuura05, Wesson08, Tsamis08}, star-forming regions
\citep{Alves2001, Thompson2002} and supernovae \citep{Sugerman06}.  In one of
the best studied PNe \object{NGC 7293 (the Helix Nebula)}, knots in the inner
regions have a comet-like shape \citep[e.g.][]{Zanstra55,Meaburn92, Odell96}
and are thus known as cometary knots.  Their apparent ubiquity in PNe has lead
to the assertion that all circumstellar nebulae are clumpy in structure
\citep{Odell02, Meixner05}.  Understanding their physical nature, such as line
excitation and formation mechanism, is essential to understanding the dominant
physics governing nebulae.

In order to discuss the nature and origin of the density inhomogeneities and
their manifestation in PNe, we first need to define our terminology. The term
``cometary knot'' is used to refer to structures that include both an
elliptical head and a trailing tail (although both these structures come in a
variety of shapes). Consequently we define ``knot'' to mean the whole
``cometary'' stucture (head and tail, if a tail exists on a given
knot-head). These structures are sometimes referred to as globules, but we
prefer to use the term ``globule'' to refer to the head only (and in some
cases there is only a head, in which case it is both a knot and a globule). In
addition to knots, various previous work on small-scale inhomogenities have
also defined FLIERS \citep[Fast Low-Ionization Emission
  Regions;][]{Balick1998} and LIS \citep[low-ionization
  structures;][]{Goncalves2001}.  Knots are a subset of LIS, but not of
FLIERS (too slow). In terms of the formation and shaping of knots, we also
define the term ``core'', which is taken to be a density enhancement in the
nebula which is subsequently molded by some mechanism into a knot.

\begin{figure}
\centering
\resizebox{\hsize}{!}{\includegraphics*[166,236][442,562]{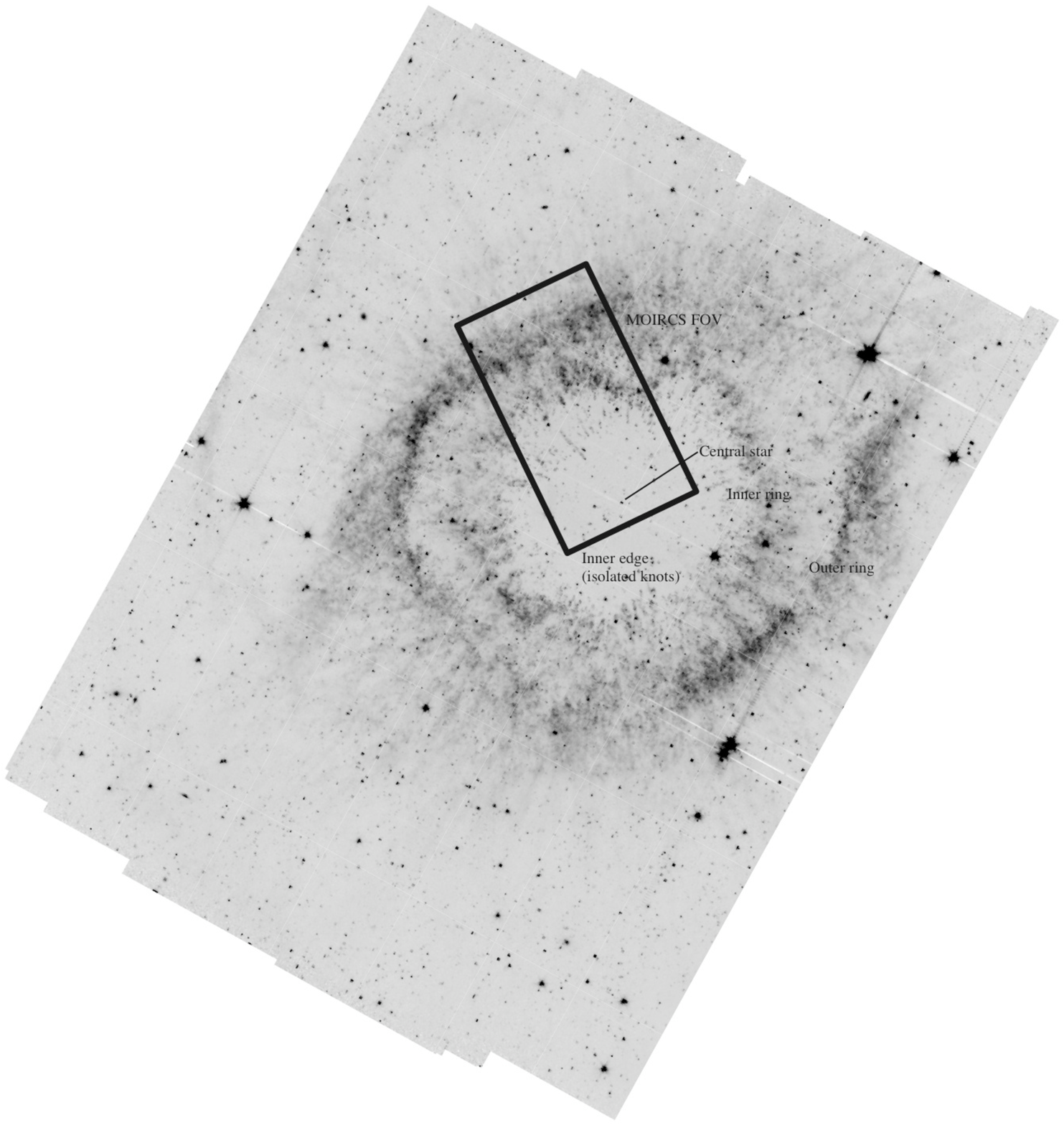}}
\caption{The region observed by the MOIRCS (4'$\times$7') is indicated by a box
on the Spitzer 3.6\,$\mu$m image \citep{Hora06}. The north is top and the east is left.
Two bright rings (inner and outer rings) are found in the 3.6\,$\mu$m image,
as well as several isolated knots in the inner edge of H$_2$ region.
MOIRCS covers approximately one-eighth of these two rings.
\label{fig-spitzer} }
\end{figure}

We have obtained a new high resolution H$_2$ image of the Helix Nebula using
MOIRCS \citep[Multi-Object Infrared Camera and Spectrograph; ][]{Ichikawa06}
on the 8.2-meter telescope, Subaru.  Two slightly elongated ellipses
(Fig.\,\ref{fig-spitzer}) of NGC\,7293 are interpreted as projected rings: an
inner ring (from about 200$''$ up to 500$''$) and an outer ring (up to
740$''$)\footnote{Note that this is only a description of the nebula's
  appearance on the sky, and is not intended to denote its 3-d stucture
  \citep[c.f.][]{Odell04}}.
Inside the inner ring, i.e., the inner edge, isolated knots have been found
\citep[e.g.][]{Odell96}.  The MOIRCS wide field of view (FOV;
4$'$$\times$7$'$) yields an image covering the inner and outer rings as well
as the central star (CS) as indicated in Fig.\,\ref{fig-spitzer}.  The
excellent seeing ($\sim$0.4$''$ in FWHM; full width at half maximum) allows
resolution of the individual knots (typically 1--3$''$ in diameter).
Combining the large FOV and high spatial resolution, we are able to study the
radial variation of knot shapes and number densities.

\section{Observations and data-reduction}

Infrared images of NGC\,7293 were taken with the MOIRCS on the Subaru
Telescope at Mauna Kea, Hawaii, USA on 2007 June 25 (UT) during the twilight.
The sky condition was photometric.  We used telescope nodding to 1\degr\ to
the east to subtract the sky background.  Jitter was used to minimize the
influence of hot pixels.  The total exposure time was 100\,s$\times$7 on
source.  Telescope ambient seeing in the optical was 0.5$''$ and the spatial
resolution of the reduced image is measured as $\sim$0.4$''$ (FWHM), measured
by stars within the field.  The airmass during the observations was 1.3.
MOIRCS is installed with two 2028$\times$2028 pixel HAWAII-2 arrays, namely
CH1 and CH2, and pixel scale is 0.117$''$.  The H$_2$ filter has a central
wavelength of 2.116\,$\mu$m and a width of 0.021\,$\mu$m (FWHM). Continuum
emission from the nebula is probably negligible at this wavelength
\citep{Matsuura07} at least at the inner edge, and the H$_2$ $v=1 \rightarrow
0$ S(1) line is dominant.  The coordinates of the center of the final image
are RA=22h29m43.02s, DEC=$-$20d47m33.3s (J2000) and the position angle is
25.4\degr\ east of north.  The MOIRCS observed region is indicated on
\citet{Hora06}'s Spitzer image (Fig.\,\ref{fig-spitzer}), which shows the
general location of the inner and outer rings and the inner edge of NGC\,7293.
The MOIRCS image covers the CS and part of both of these two rings.

Our Helix image has an angular resolution comparable to previous HST NICMOS
images \citep{Meixner05, Odell07}.  The pixel scale of NICMOS NIC3 camera is
0.2\,$''$, the Nyquist sampling provides a resolution (FWHM) of
0.4\,$''$. \citet{Meixner05} and \citet{Odell07} have taken two frames at each
field with slightly different positions to increase the spatial
sampling. Depending on the pointing accuracy, the final images combining two
different positions would create a spatial resolution of 0.2--0.4\,$''$. Indeed,
\citet{Odell07} reported the FWHM of a star of 0.42\,$''$.  We measured FWHMs
of two stars in two different fields in \citet{Meixner05}, yielding
0.20--0.42\,$''$.  Our MOIRCS (0.4\,$''$) image is among the highest angular
resolution images of multiple knots of the Helix at 2\,$\mu$m, after the
observations of a single knot at 0.05\,$''$ pixel scale of \citet{Matsuura07}.

We use {\it IRAF} and the {\it IRAF}-based software {\it QMCS} which has been
developed by one of the authors (I.T.)  for the data reduction.  We first
adopted the correction of the flat field, the sky subtraction and the
distortion correction, and a mosaic image was created from seven target frames
by registering field stars and by adopting masks on bad pixels.  In the final
image, there are low level artifacts in the form of multiple circular rings on
the image taken with one of the arrays. This pattern is caused by interference
of the filter.  In the CH1 image (the south part of the image), this pattern
was reduced using defringe$_{-}$moircs.pro, which is a part of `SIMPLE Imaging
and Mosaicing Pipeline' (Wang, in preparation).  The sensitivity is
approximately $3\times10^{-19}$\,ergs\,s$^{-1}$\,cm$^{-2}$\,\AA$^{-1}$ at a
signal-to-noise ratio of 5.  The RMS noise on the sky level was measured over
five hundreds circular apertures (radius of 0.39\,$''$; approximately twice
the FWHM of the seeing limit) on the blank space of the CH1 image.  The median
of the RMS noise distribution is taken as the sensitivity.  We use 2MASS
magnitudes of three stars in CH1, and we assume that their $Ks$ magnitudes are
identical in the H2 filter.  No such measurement is possible in CH2, because
of the high density of knots.  A similar measurement is applied to NICMOS
images \citep{Meixner05}, giving
$3\times10^{-18}$\,ergs\,s$^{-1}$\,cm$^{-2}$\,\AA$^{-1}$. The sensitivity of
our MOIRCS image is better than the NICMOS images by an order of magnitude.

\section{Results} \label{sect-results}

\subsection{Overall descriptions of the H$_2$ image}

\begin{figure*}
\centering
\resizebox{\hsize}{!}{\includegraphics*{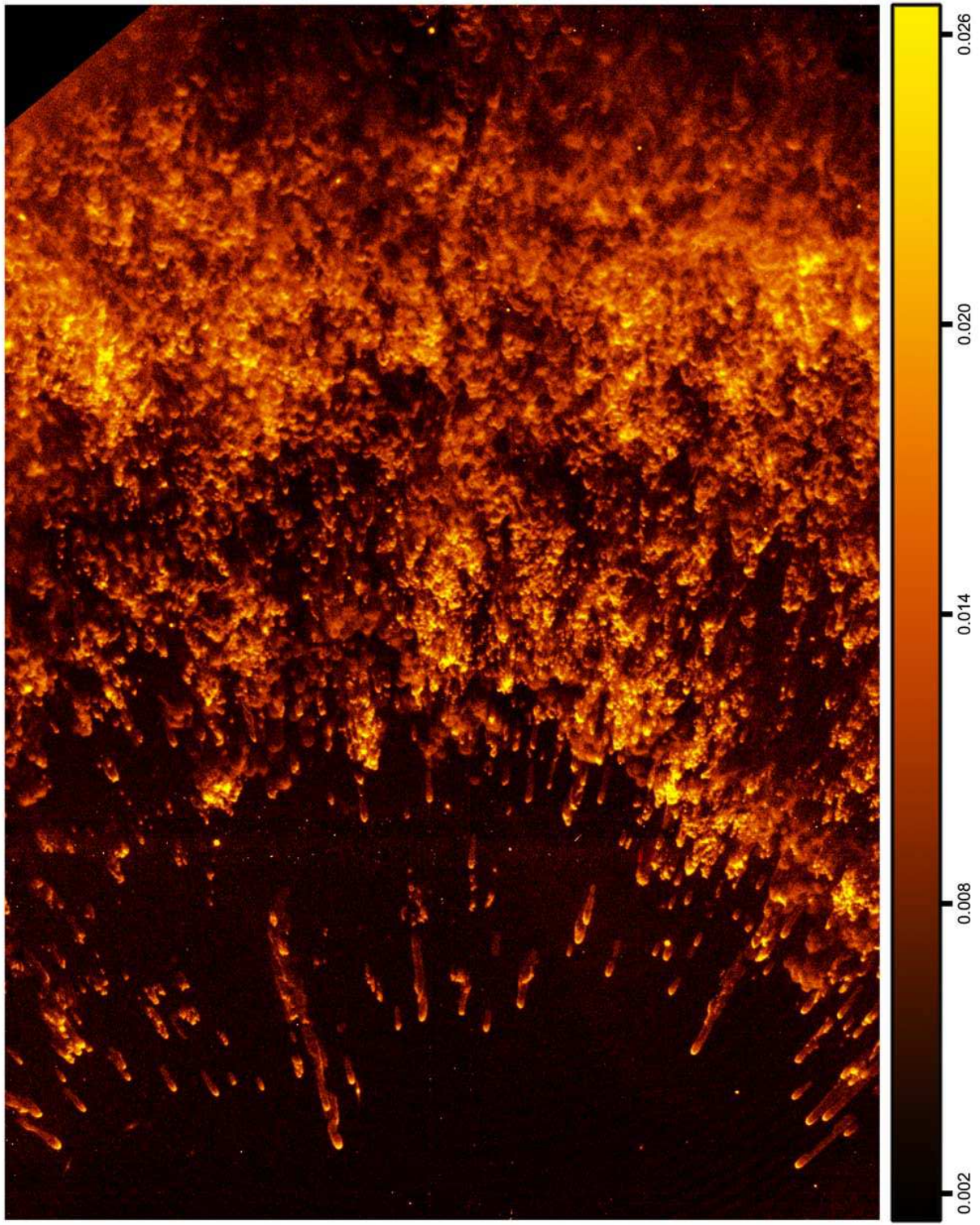}}
\caption{
Part of the MOIRCS H$_2$ image covering 
about three-quarter (5.10'$\times$3.45') of the obtained image.
The CS is outside of this image towards the bottom.
Isolated cometary knots are found near the CS.
Towards top of this figure, knots are more clustered. 
Here knots are not well-resolved, and rather seen as clumps.
The color scale is relative intensity to the maximum counts of the image.
\label{fig-enlarge} }
\end{figure*}

\begin{figure}
\centering
\resizebox{\hsize}{!}{\includegraphics[8,248][569,619]{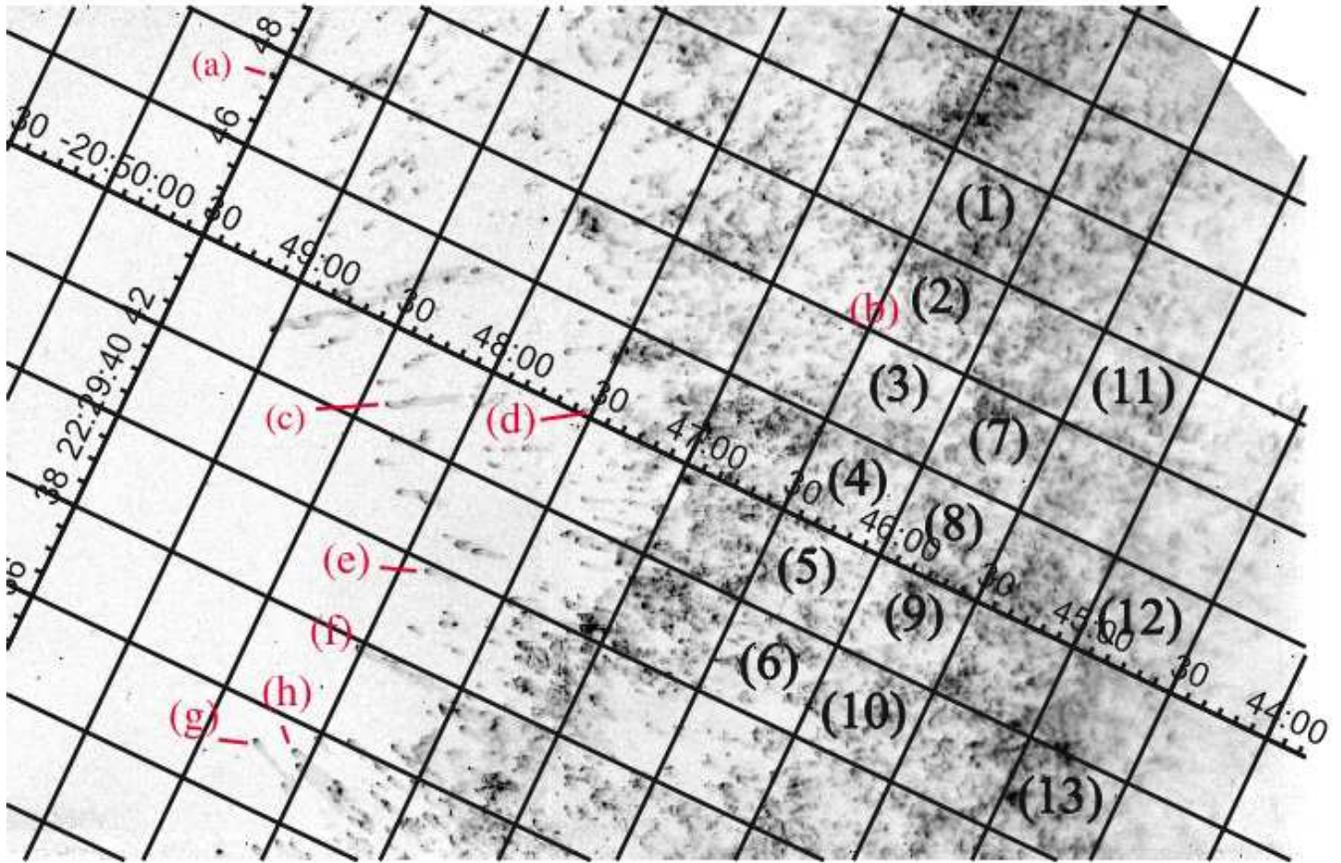}}
\caption{
The locations where the knots enlarged in Fig.\,\ref{fig-indivisual} are
found,
and where knot densities are counted, shown in
close up in Fig.\,\ref{fig-examples}.
\label{fig-regions} }
\end{figure}

\begin{figure}
\centering
\resizebox{0.8\hsize}{!}{\includegraphics*[20,215][465,695]{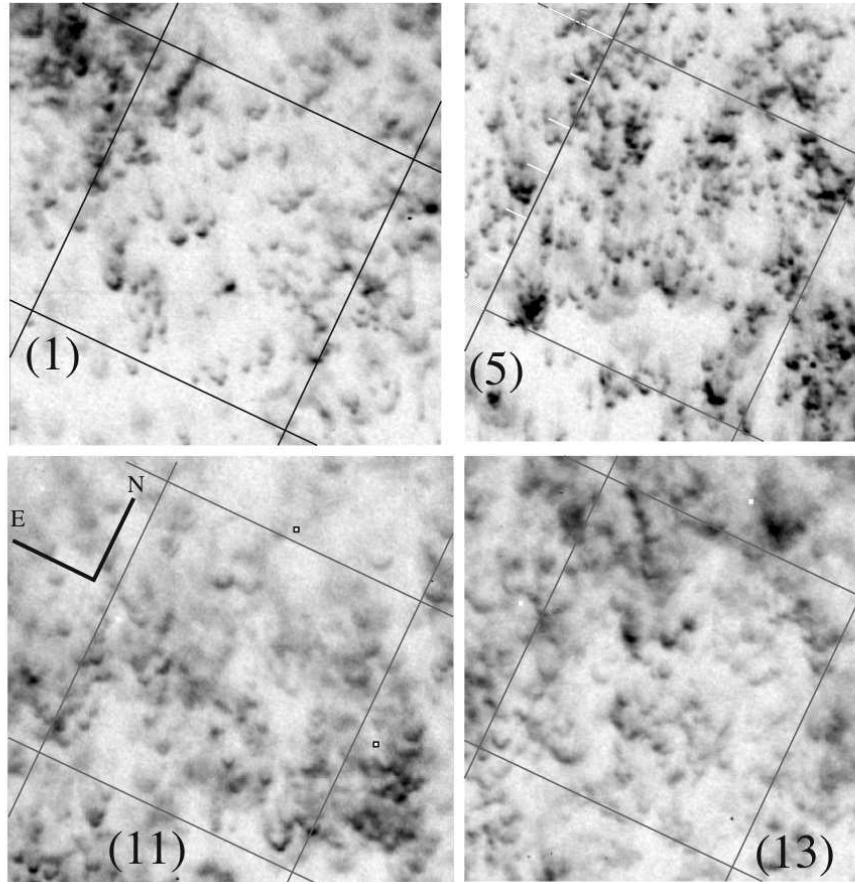}}
\caption{
Close up of parts of the inner ring (1 \& 5) and the outer ring (11 \& 13)  in Fig.\,\ref{fig-regions},
where knot densities are counted. One grid is 30$''$ by 30$''$.
Clear crescent knots with tails are found in the inner ring, while 
fainter knots and diffuse/unresolved components are found in the outer ring.
\label{fig-examples} }
\end{figure}

 Fig.\,\ref{fig-enlarge} shows the MOIRCS image of NGC\,7293 in the $v
 \rightarrow $1--0 S(1) H$_2$ line, while Fig.\,\ref{fig-regions} shows a grid
 reference indicating the R.A. and Dec. ($J$2000) overlaid on the image,
 together with a numbered grid to guide the reader to which region we are
 discussing.  Clumps of H$_2$ gas are found throughout the entire image
 (except the area close to the central star), supporting \citet{Meixner05} and
 \citet{Hora06}.  The knots are isolated at the inner edge (distance from the
 CS $r<$3.5$'$) and these knots are accompanied by tails pointing directly
 away from the CS.  The areas enlarged in Fig.\,\ref{fig-examples} are
 indicated in Fig.\,\ref{fig-regions}. Individual knots and their tails remain
 resolved throughout the inner ring (3.5$'$$<r<$4.5$'$).  While the H$_2$
 emission is clumpy in the outer ring (4.5$'$$<r<$6.4$'$), the structures are
 different; individual knots are not as easy to distinguish, and tails do not
 appear to be collimated into the radial direction (Fig.\,\ref{fig-examples}).
 The outer ring is brighter than the inner ring in H$_2$, as found by
 \citet{Speck02}.  Our MOIRCS image clearly shows that the morphology of the
 knots changes from the inner edge to the outer ring.
In general, the well-resolved knots in the inner edge and inner ring show a 
`cometary' shape, i.e.  an elliptical head including a bright
crescent-shaped tip and a tail.  In the inner region, the crescents always
face the CS, and the tails stretch in the opposite direction.  Examples of
morphologies are found in Figs.\,\ref{fig-examples} and \ref{fig-indivisual}.
Such knots are found up to $r\sim$4.5$'$ (approximately the lower half of
Fig.\,\ref{fig-enlarge}) from the CS, i.e., the inner ring.  From previous
observations, the cometary morphology has only been recognized at the inner
edge \citep{Huggins02, Hora06, Matsuura07}.  Our MOIRCS image shows that these
morphologies are found throughout the inner ring.

Beyond $r\sim4.5'$ (the outer ring), tails are not always obvious; some knots
appears to have crescents only (Fig.\,\ref{fig-examples}), as described 
by \citet{Meixner05} and \citet{Odell07}.

\subsection{Detailed descriptions of tails}

The shapes of the tails are best studied in H$_2$ due to its brightness.
\citet{Matsuura07} studied a knot at the inner edge of the H$_2$ emitting
region and showed that the optical [N{\small II}] +H$\alpha$ intensity falls
to less than 10\,\% at 1$''$ away from the brightest tip, while the H$_2$
intensity decreases more gradually to 10\,\% at 5$''$.

\begin{figure*}
\centering
\resizebox{0.9\hsize}{!}{\includegraphics*{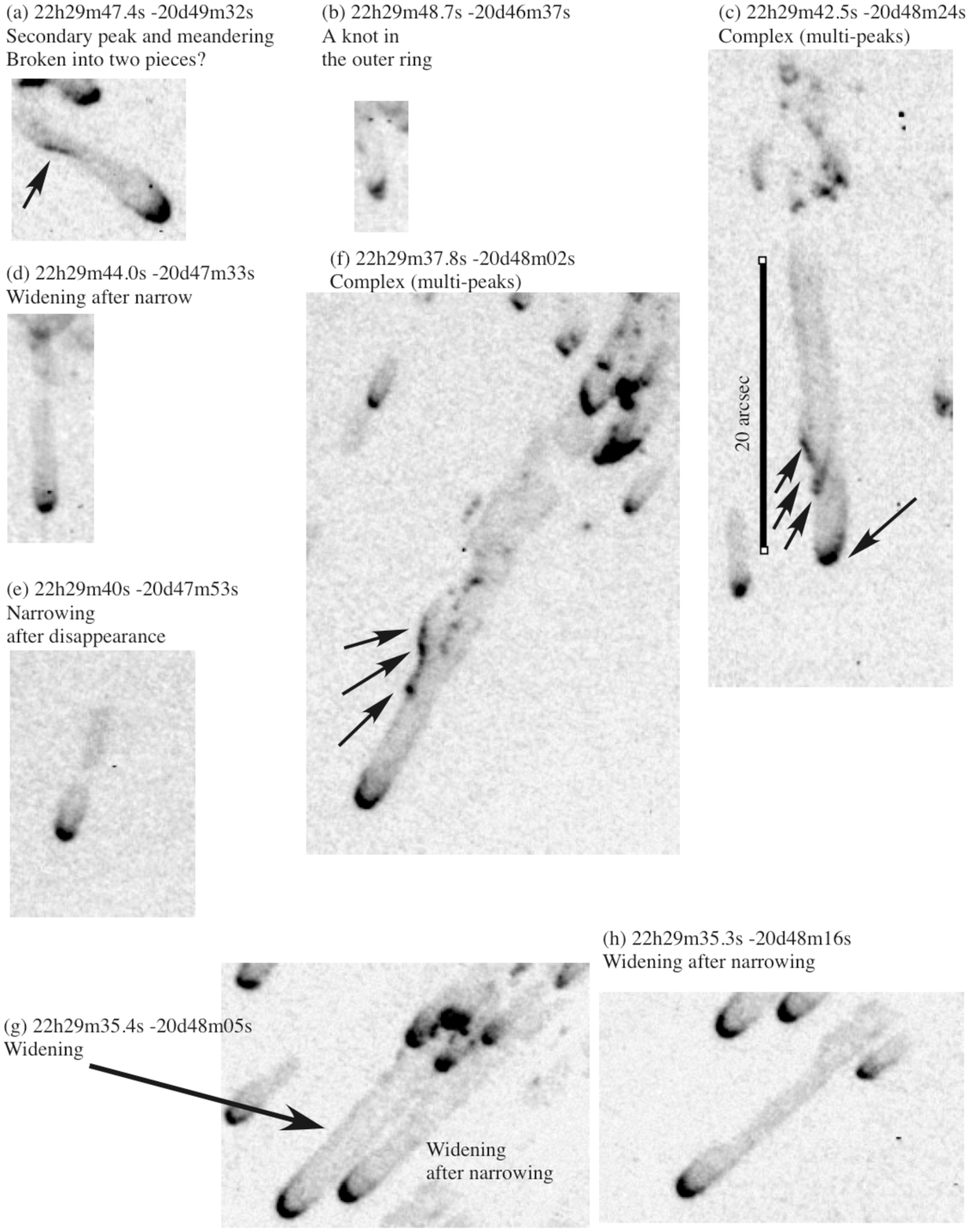}}
\caption{Various shapes of knots.
 Coordinates are in J2000. The knot (c) is identical to 44 in \citet{Meaburn98} and
(f) is identical to 1 in \citet{Meaburn98} and  378-801 in \citet{Odell05} and also found in
\citet{Huggins02} and \citet{Odell07}.
\label{fig-indivisual} }
\end{figure*}

Fig.\,\ref{fig-indivisual} displays a variety in tail shapes.  The knots in
this figure are mostly located in the inner edge (Fig.\,\ref{fig-regions}) and
they are well isolated.  The most common shape is a narrowing tail, which was
also found for an individual knot at the highest angular resolution (0.05$''$
per pixel) \citep{Matsuura07, Matsuura08}. However, some tails first narrow
then widen with distance from the bright tip, such as knot (h) in
Fig.\,\ref{fig-indivisual}.  Similarly, some tails show a brightness gap at
the midpoint along the tail (d) and (e).  Such tail shapes have not been
reported previously.  There is one case of a continuously widening tail
straight from the crescent (g).  Knots occasionally show complex structures,
with a secondary peak or sometimes multiple intensity peaks in the tail, often
accompanied by a meandering tail (a), (c), and (f).
%
%
The orientation of the tail often changes at the secondary peak (a) and (f).
As the secondary peak is aligned with the tail, and no crescent is found
nearby, the secondary peak is physically associated with the primary peak.
The wide variety of tails may be linked with the formation mechanisms of knots
and tails, or they trace an interaction with the local environment. This will
be discussed in Sect.\,\ref{shaping}.

\subsection{Number density of knots and H$_2$ surface brightness}

\begin{figure}
\centering
\resizebox{\hsize}{!}{\includegraphics*[0,270][525,569]{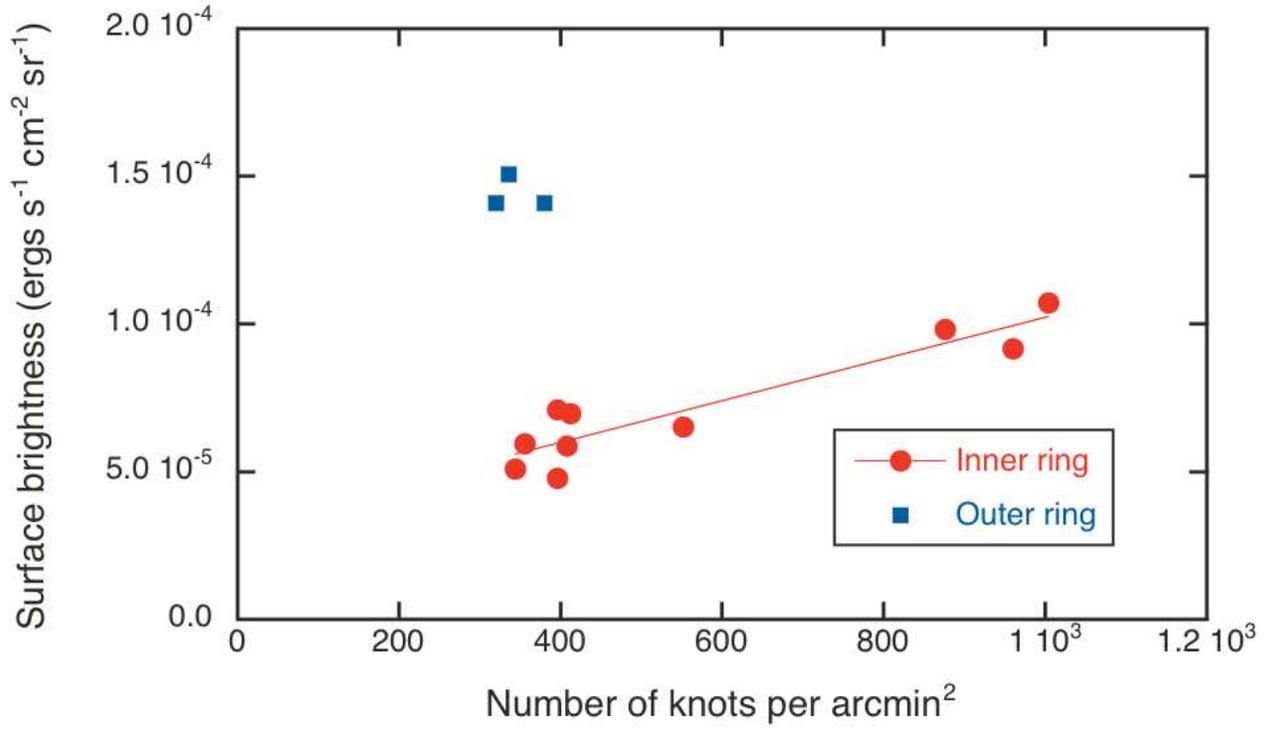}}
\caption{
The number density of knots in 30$''$$\times$30$''$ boxes counted in MOIRCS image,
and averaged surface brightness within these boxes calculated
from \citet{Speck02}.
The number density of knots correlates well with the surface brightness in 
the inner ring.
The line shows a least square fit to the data of the inner ring.
The Ssurface brightness in the outer ring
is approximately twice as high as  in the inner ring
for a given number density of knots. 
\label{fig-count} }
\end{figure}

We counted the number density of knots and compared with the H$_2$ surface
brightness.  Fig.\,\ref{fig-count} shows the number of knots counted in
1$\times$1\,arcmin$^2$ regions, as a function of average surface
brightness within these regions calculated from \citet{Speck02}.  
In the inner ring, the surface brightness of H$_2$ correlates well with the
number density of H$_2$ knots, rather than with the brightness of individual
knots, or the distance from the CS.

For a similar number density of knots, the outer ring has twice as high
surface brightness as does the inner ring.  This is due to a contribution from
blended H$_2$ tails, which occupy a larger area than the knot-heads in the
outer ring, as found in Figs.\,\ref{fig-enlarge} and \ref{fig-examples}.
On average, the H$_2$ surface brightness from the outer ring is higher
than that from the inner ring.

In order to estimate the total number of knots in this nebula, we divide the
nebula into three zones, the innermost edge ($r$=2.2--3.5$'$), the inner ring
($r$=3.5--4.5$'$) and the outer ring ($r$=4.5--6.4$'$).  From the MOIRCS
image, we estimate the knot number densities of these regions to be 20, 500
and 400 arcmin$^{-2}$, respectively.  We assume that knots are uniformly
distributed in these circular rings.  The total number of knots is estimated
to be $4\times10^4$ in the entire nebula.  This is a factor of 1.7 larger than
the estimate by \citet{Meixner05}.  The difference is caused by the assumed
knot number densities in the inner ring; \citet{Meixner05} used a lower number
density (148 arcmin$^{-2}$) throughout the entire nebula based on observations
of the outer ring.  We counted knot number densities in relatively less
crowded regions within our image so as to minimize confusion problems.
Despite that, these regions are bright in the Spitzer 3.6\,$\mu$m image, where
the 3.6\,$\mu$m band also represents H$_2$ emission
\citep[Fig.\ref{fig-spitzer}; ][]{Hora06}.  \citet{Meixner05} counted much
fainter regions, where the density is lower.
The total number of knots provides an estimate of the total molecular and
neutral hydrogen gas mass.  Adopting a knot mass of
1.5\,$\times10^{-5}$\,M$_{\sun}$ \citep{Odell96}, the total molecular mass is
0.6\,M$_{\sun}$ in this nebula.  This estimate is a factor six larger than
the molecular mass estimated from CO observations \citep{Young99}.
The difference could be due to the large fluctuations of knot mass.  Since the
 ionized gas mass is estimated as 0.3\,M$_{\sun}$ in this nebula
\citep{Henry99}, a substantial amount of molecular gas remains.  Most of the
H$_2$ gas is found in the outer ring, thus the distance from the CS, i.e. UV
radiation strength is a key.  Only knots with large optical depth could
survive in the inner ring.

\subsection{Comparisons with optical images}\label{sect-optical}

\begin{figure}
\centering
\rotatebox{270}{ 
\begin{minipage} {10cm}
\resizebox{\hsize}{!}{\includegraphics*[55, 35][500, 807]{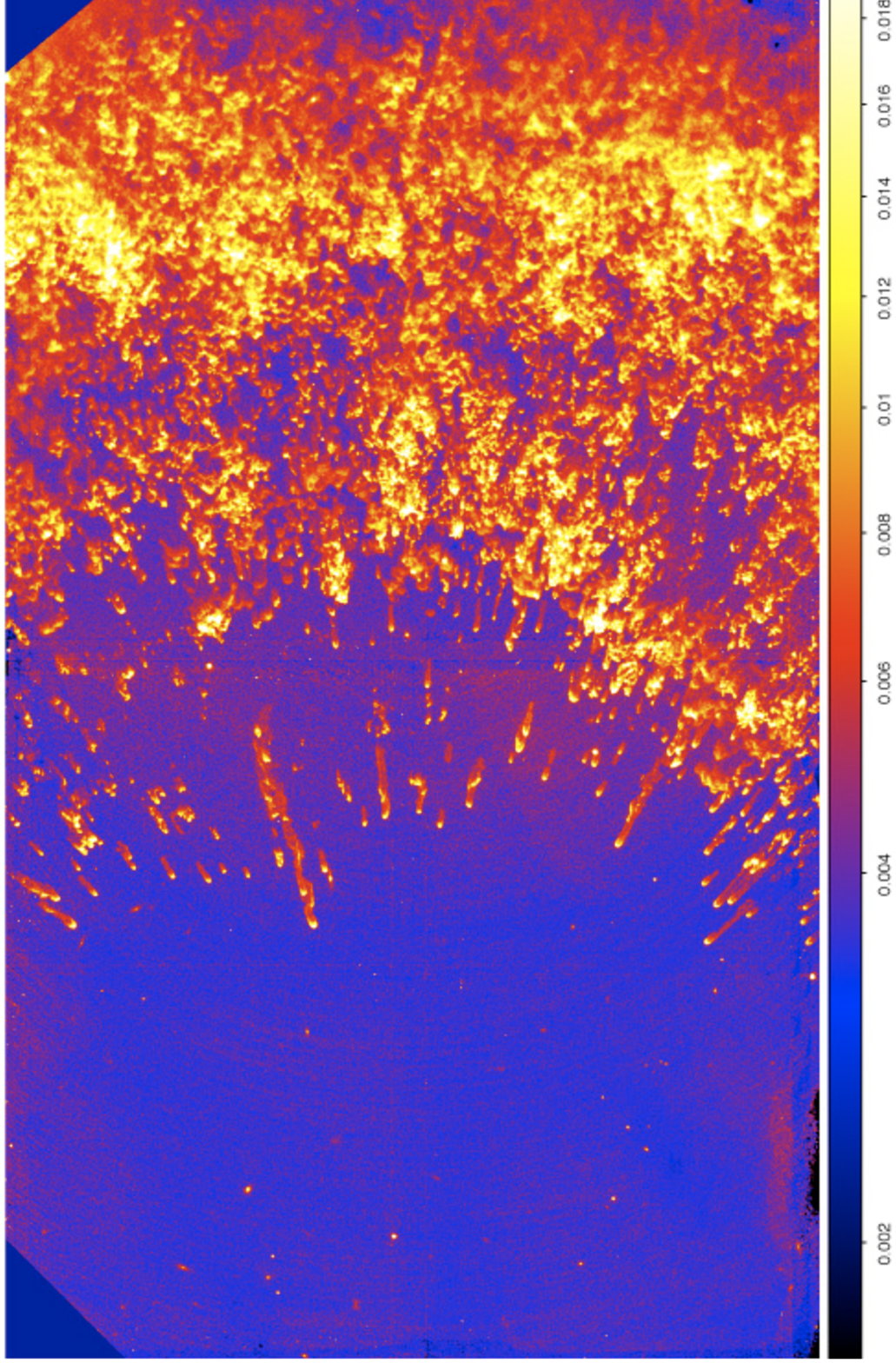}}
\end{minipage}}
\rotatebox{270}{ 
\begin{minipage} {10cm}
\resizebox{\hsize}{!}{\includegraphics*[55, 35][500, 807]{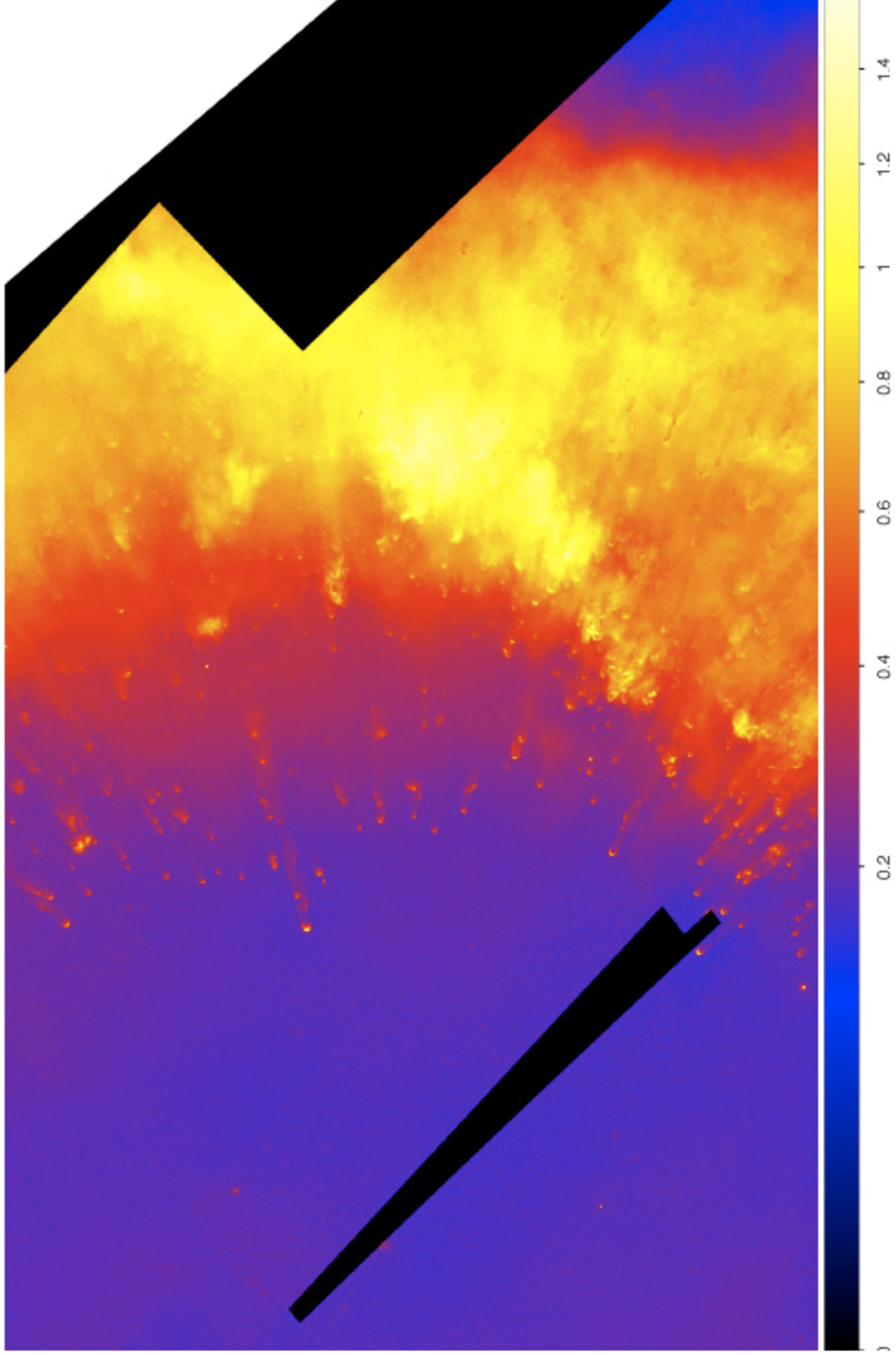}}
\end{minipage}}
\caption{MOIRCS 7'$\times$4'
H$_2$ image (top) and corresponding region
of the HST F658N ([N{\scriptsize{II}}] +H$\alpha$) image \citep{Odell04}.
\label{fig-HST} }
\end{figure}

\begin{figure*}
\centering
\resizebox{\hsize}{!}{\includegraphics*[10, 265][420, 620]{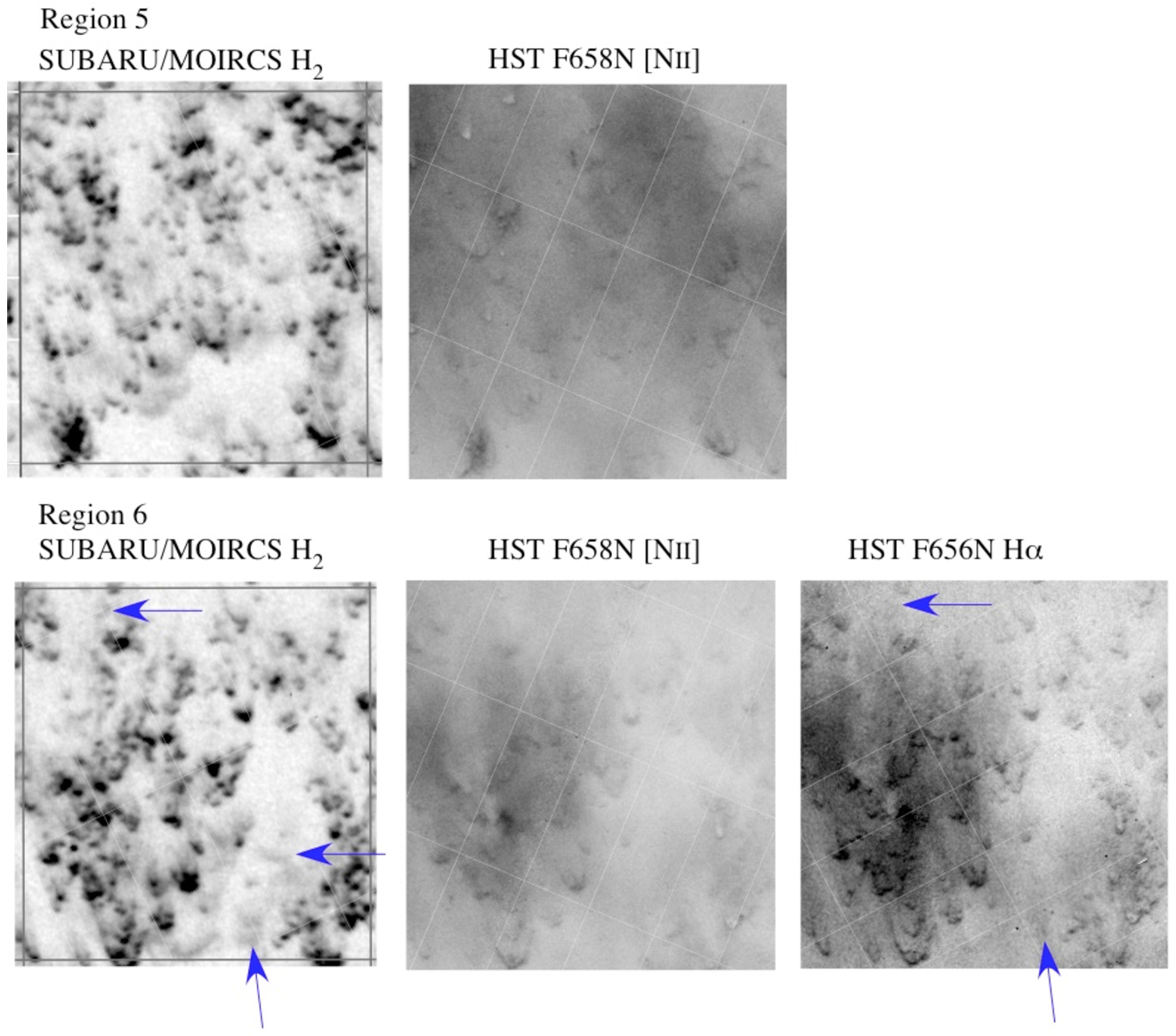}}
\caption{Enlarged images of regions 5 and 6 in 2.12\,$\mu$m H$_2$, F658N (658\,nm;
mainly [N{\scriptsize II}] with some contamination of H$\alpha$) 
and F656N (656\,nm; H$\alpha$ with some contamination of [N{\scriptsize II}]).
\label{fig-MOIRCS-HST} }
\end{figure*}

\begin{figure*}
\centering
\resizebox{\hsize}{!}{\includegraphics*[5, 230][482, 666]{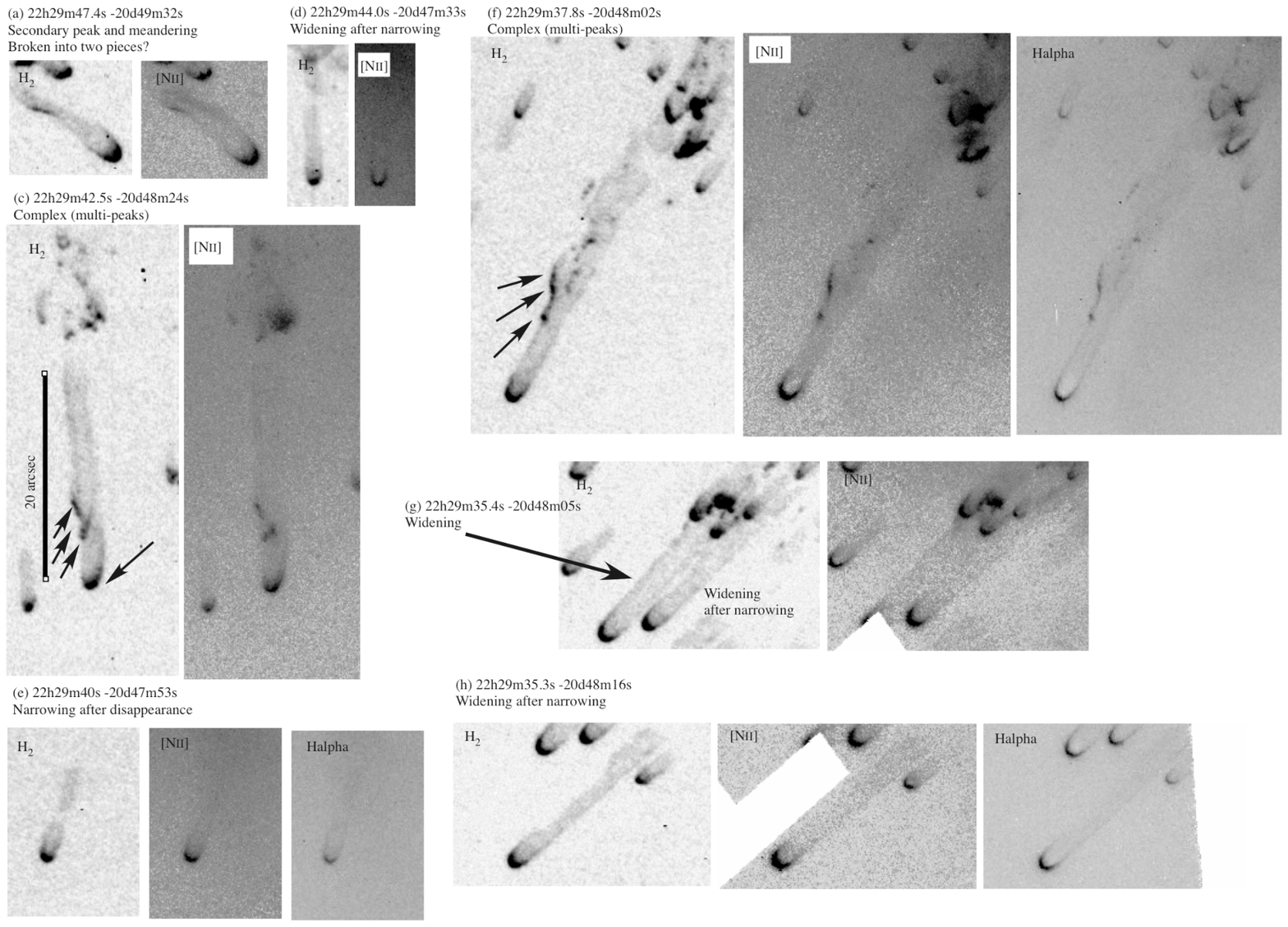}}
\caption{Enlarged images of knots in 2.12\,$\mu$m H$_2$, F658N (658\,nm;
mainly [N{\tiny II}] with some contamination of H$\alpha$) 
and F656N (656\,nm; H$\alpha$ with some contamination of [N{\tiny II}]).
\label{fig-MOIRCS-HST-knots} }
\end{figure*}

NGC\,7293 has been targeted by various optical observations.
Figs.\,\ref{fig-HST}, \ref{fig-MOIRCS-HST} and \ref{fig-MOIRCS-HST-knots}
shows a comparison of the MOIRCS H$_2$ image with HST images in the F658N and
F656N filters \citep[][respectively]{Odell05,Odell96}.  The F656N data at 656
nm are dominated by H$\alpha$ with some contamination from [N{\small II}],
whereas the F658N data at 658 nm are dominated by [N{\small II}], with
negligible contamination from H$\alpha$.  The F656N image has been taken only
for smaller parts of the Helix Nebula.

Fig.\,\ref{fig-HST} shows the overall images of H$_2$ and F658N.  The
brightest places in H$_2$ and [N{\small II}] are found in different locations
within the nebula.  The inner ring is brighter than the outer ring in
[N{\small II}] whereas the outer ring is brighter in H$_2$.  All knots in the
optical image have counterparts in H$_2$ emission.  Some knots cause
extinction in the optical (and appear dark in the image), as found near the
bottom of the [N{\small II}] image (Fig.\,\ref{fig-HST}).  These knots are
located in the foreground of diffuse ionized emission, and all of these
absorption knots in [N{\small II}] have emission counterparts in the H$_2$
image \citep[c.f. the case of \object{NGC\,6720}; ][]{Speck03}.  In contrast,
not all H$_2$ knots have corresponding knots in the [N{\small II}] image,
particularly in the inner ring where [N{\small II}], emission from the
inter-knot region overwhelms the emission from knots, if any.
\citet{Meixner05} and \citet{Odell07} suggested that H$_2$ knots always have
some counterparts in [N{\small II}] but this is only the case in the outer
ring and at the inner edge.  In the outer ring, diffuse [N{\small II}]
components are relatively weak, and in the inner edge, all of the knots are
isolated.  This contrast can be explained if NGC\,7293 takes the morphology of
a pole-on bipolar nebula projected onto two rings \citep{Odell04, Meaburn05}.
Moreover, these rings are quite thick and contain layers of knots and diffuse
components within.

Fig.\,\ref{fig-MOIRCS-HST} shows a close-up comparison of the H$_2$, [N{\small
    II}] and H$\alpha$ images in the inner ring, labeled as regions 5 and 6 in
Fig.\,\ref{fig-regions}.
The F656N image is available only for region 6 within the MOIRCS 13
regions defined in Fig.\,\ref{fig-regions}.
Knots are clearly found in the H$_2$ images for both regions 5 and 6, whereas
less than one-third of those crescents/heads are recognised in [N{\small II}]
image.
%
In the [N{\small II}] images, there is little to no structure defining the
knots, with the ionized emission being more diffuse.
In region 6, the H$_2$ crescents have counterparts in H$\alpha$, except where
indicated by arrows in Fig.\,\ref{fig-MOIRCS-HST}.

H$_2$ is associated only with high density knots, whereas the ionized gas is
located both at the surface of the knots and in the inter-knot space. The high
density of the knots leads to high brightness in the knot locations for H$_2$
and to a lesser extent, H$\alpha$. The H$\alpha$ contrast is reduced by the
inter-knot component.

We further compare H$_2$, [N{\small II}] and H$\alpha$ images of well-isolated
knots in the inner edge of the nebula (Fig.\,\ref{fig-MOIRCS-HST-knots}).
These images are aligned such that the tips of the knots are coincident at
these three wavelength.  \citet{Odell05} reported a displacement of the
[N{\small II}] tips with respect to those in NICMOS H$\alpha$ and H$_2$ by
$\sim$0.1$''$: this displacement is minor compared to our 0.4$''$ resolution.
All knots except (b) were detected in [N{\small II}], but only three were
found in H$\alpha$.  All crescents are found in all three wavelengths.  The
secondary peaks in knots (a), (c), and (f) are also found in all wavelength
images: fainter local peaks in the H$_2$ image of knot (f) can be seen in the
H$\alpha$ image, but not in the [N{\small II}] image.  In the [N{\small II}]
and H$\alpha$ images, high emission from the inter-knot region is found.  Due
to this high inter-knot emission and faintness of tail emission, the narrowing
of the tails is not always clearly seen in the [N{\small II}] image.

\section{Discussion}

\subsection{Origin of the molecular hydrogen in knots}

Our observations show that at the inner edge and the adjacent regions of the
nebula, the H$_2$ is found only in knots. In the outer ring there is a faint
dispersed H$_2$ component which can be interpreted as overlapping blended
tails.  Therefore, we conclude that the H$_2$ is associated solely with the
knots, following \citet{Speck02} and \citet{Meixner05}.  There are two
plausible explanations for such an association: (i) the large optical depth of
the knots protects previously formed H$_2$ from the harsh UV radiation, or
(ii) the H$_2$ formed in-situ in the knots, where high densities and low
temperatures allow for molecule formation.

The choice between these two explanations is linked to the origin of the
knots: did these form at late times, after ionization of the planetary nebulae
began \citep[e.g.][]{GarciaSegura2006}, or did they originate in a molecular
wind \citep[e.g.][]{Redman03}? In the first case, H$_2$ must have formed
in-situ, while in the second case the molecular content may have survived from
the AGB wind.

In order to test whether H$_2$ formed during the AGB phase could have survived
into the PN phase, we have investigated the lifetime of molecular hydrogen.
Using the dissociation rate of H$_2$, including self-shielding \citep{Draine96}
%
%
%
%
and $G_0=8$ and $n_{\rm H_2} = 8\times10^4$\,cm$^{-3}$ \citep{Matsuura07}, 
the time scale to dissociate H$_2$ is estimated to be $8\times10^7$\,years.
Thus, pre-existing H$_2$ can survive in dense regions within the current
conditions.  However, the UV radiation field will have been much stronger in
the past. The star is currently on the cooling track, and is much less
luminous than during its earlier post-AGB evolution.
If $G_0=10^3$, the dissociation time scale is $8\times10^4$\,years.  The
high-luminosity regime phase for relatively massive stars lasts for less than
10,000\,yrs \citep{Zijlstra2008}.  Hence, the H$_2$ associated with knots may
have survived such an intense UV radiation field for this early post-AGB
period.

%


In order to investigate whether H$_2$ can be formed in-situ in PNe,
\citet{Aleman2004} calculated H$_2$ formation in (partly) ionised gas.  The
reactions involved include H$^-$\,+\,H and H$_2^+$\,+\,H.  The H$_2$ formation
rate in the ionized gas is estimated to be 10$^{-10.5}$\,cm$^{-3}$s$^{-1}$ in
total, which is slightly larger than the H$_2$ destruction rate.  However, the
fraction of H$_2$ with respect to H is extremely small ($10^{-5}$).
\citet{Aleman2004} concluded that some H$_2$ could survive the ionized state,
but it is not yet clear if H$_2$ can be formed from the ionized gas during the
PN phase.

The densities within the knots are estimated to be $10^6\,\rm cm^{-3}$
\citep{Meaburn98} to $3 \times 10^5\,\rm cm^{-3}$ \citep{Huggins02}. At
$10^6\,\rm cm^{-3}$, H$_2$ formation on dust grains becomes significant; the
time scale is of the order of $10^3\,\rm yr$.  Thus, in the densest knots the
formation can proceed faster than the age of the nebula.  However, if the
densities are lower, the required life time of the knots becomes similar to the
age of the nebula. In such a case, the knots pre-date the ionization, removing
the need to (re-)form the H$_2$ later.

%
%
%
%

In summary, the time scale suggest that part of the H$_2$ is primordial,
i.e. formed during the AGB phase, and survived the ionization of the nebula.
In  dense knots, (re-)formation of H$_2$ is possible.

\subsection{Knot formation models}

There have been various suggested mechanisms for knot formation. The two main
scenarios are: (1) the knots are primordial density enhancements, i.e. they
started their formation during the AGB phase
\citep[e.g.][]{Zanstra55,Dyson1989,Soker1998}\footnote{At the time of
  \citet{Zanstra55}, the AGB phase was not established yet, but we count that
  the phase before gas ionization mean AGB phase and possibly post-AGB phase};
(2) the knots form in situ, as a result of the onset of the PN phase
\citep[e.g.][]{GarciaSegura2006, Capriotti1971}.

The in-situ formation mechanisms can be further divided into subcategories:
(i) turbulent (Rayleigh-Taylor; RT) instabilities as a result of interaction
between the slow and fast winds \citep[e.g.][]{Mathews1969, Vishniac1994,
  Dwarkadas98, GarciaSegura2006}; (ii) turbulent (Rayleigh-Taylor; RT)
instabilities at the ionization front \citep[e.g.][]{Capriotti1971}; (iii)
shadowing effects \citep[e.g.][]{VanBlerkom, Soker1998}; and (iv) radiative
shocks \citep[e.g.][]{Walder}. \citet{Capriotti1973} used his model to predict
the mass ($10^{-6}$\,M$_\odot$) and number of knots ($\sim 10^5$) in a nebula,
consistent with current observations.

A pre-ionization origin of the knots appears to be plausible. The system of
knots has an expansion velocity which is typical for AGB winds but is less
than that of the ionized gas \citep{Meaburn98}. The last point suggests that
the knots formed after the AGB wind reached its terminal velocity (at $\sim
10$\,AU from the star), but before the overpressure from the onset of
ionization caused further acceleration \citep[e.g.][]{Ueta06}.  The fact that
a very high fraction of the total mass is found in the knots, also is more
easily reconciled with an early origin.

To form knots in an expanding wind requires compression of the gas, within an
overpressured (heated) surrounding region. This may happen at the ionization
front, as discussed above. Because the ionization front moves outward, the
inner sets of knots would be older than the outer knots, in this model. The
shock front between the fast and slow wind is another option but this front
co-moves with the expanding nebula, so that one might expect the knots to be
found only near the inner edge of the dense nebula.  This interface is located
within the ionized region, so that one may expect the knots to share the
velocity with the ionized gas. This model cannot be entirely excluded but may
be less in agreement with the observations. In this case, the H$_2$ would have
to form from ionized gas. In either of these cases, the knots would be a very
common phenomenon in planetary nebulae.

To form the knots much earlier, in the AGB wind, would require a denser region
than normally expected in AGB winds.  However, if the AGB star has a close
binary companion, a spiral shock will form in the AGB wind behind the
companion \citep{Edgar08}. This shock causes a heated, compressed region,
which a large fraction of the AGB ejecta pass through. This could provide a
possible location for the origin of the knots. In the absence of detailed
hydrodynamical models, this must remain speculative. In this model, the outer
knots are  older than the inner knots.

\subsection{Shaping knots and tails} \label{shaping}

The problem of distinguishing between formation mechanisms becomes even more
difficult once we consider the shaping of the knot heads and tails. Most
recent studies of knot shapes assume a pre-existing density enhancement and do
not consider whether it is primordial or formed during the PN
phase. Consequently, observations of knot (head and tail) shapes, may be able
to distinguish between shaping models, but are unlikely to reveal the origin
of the original density enhancement.

The shaping mechanisms can be also classified into three basic categories, with
some cross-fertilization: (a) Hydrodynamic interaction between winds and
density enhancements \citep[e.g.][]{Zanstra55,Vishniac1994,Pittard05,Dyson06,
  Pittard08}; (b) Interaction (including photoevaporation) of material with
ionizing (UV) radiation  \citep[e.g.][]{Mellema1998, Williams2003,
  GarciaSegura2006, Lopez-Martin01}; (c) The effect of shadowing by density
enhancements \citep[e.g.][]{Soker1998, Canto98}.  
These mechanisms may act together, and disentangling which mechanisms are
mutually consistent and which are mutually inconsistent is a challenge.  

In the following, we will first discuss streaming models, followed by
photo-evaporation and shadowing models. A combination is proposed based on the
observational constraints.

%

In all cases discuss below we use the term ``core'' to describe the density
enhancement which is acted upon by the various mechanisms to create cometary
knot shapes (heads and tails).

The first proposed mechanism was the interaction of a core with a wind
\citep[which may emanate from the CS or be caused by expansion of the
  inter-knot gas; ][]{Zanstra55}.  This basic concept was also the inspiration
for \citet{Dwarkadas98}, who showed that by including an evolution in the fast
wind, small clumpy structures could be formed, but the resulting shapes are
not described.
A more recent attempt to modify this concept includes both an ambient
wind, which can be sub- or supersonic, and a flow from the core
itself, where flow is assumed to have a constant ablation.  We call this a
`stream' model. The wind is due to the differential velocity between the
diffuse gas and the knots: there is no evidence for a hot wind from the
central star.
%

The interaction between the core and the wind can create both the heads and
the tails \citep{Pittard05, Dyson06}.  In this model, the crescent tip is a
bow-shock (ignoring photo-ionization) and ram pressure creates a narrowing
tail for subsonic ambient winds \citep{Pittard05} and a widening tail for
supersonic ambient winds.
%


%

The second option, i.e. interaction of material with ionizing (UV) radiation,
considers photo-ionization and photo-evaporation models, ignoring streaming
motions \citep[e.g.][]{Mellema1998,GarciaSegura2006,Lopez-Martin01}. In this
case, the surface of a core facing the CS is illuminated by UV radiation,
which causes heating and photo-evaporation from the core, and creates the
crescent tip.  At this moment, there is no detailed hydrodynamical model
present to reproduce narrowing, widening and meandering tails, using
photo-evaporation.  \citet{Williams2003} calculated ionization front
  instabilities, and obtained a jet shape. No clear resemblance is found yet
  between the modelled shape and the knots observed in the Helix Nebula,
  however, the explored parameter space was limited.

 The final possibility is the shadowing  by density 
enhancement \citep{Canto98}, leading to non-ionized regions in the shadow.  
Here, the knots
can trace a large core, consisting of neutral gas and dust, with a bright
sunny side and a dark shadow giving rise to the tail.  UV photons from the CS
do not reach the shadowed side directly, but diffuse UV radiation diffusion
illuminates the tail. To explain narrowing tail \citep{Canto98} and
widening tail \citep{Odell05}, varying angles of illumination are used.
%
%


The tails can be explained either with streaming motion, by interaction at the
ionization front or by shadowing.  Here the streaming models or ionization
models appear better able to fit the available data.  The bright (and
well-isolated) inner knot located in our field, shows clear acceleration in
[N\,{\sc ii}] along the tail \citep{Meaburn98}, by about 10\,km\,s$^{-1}$.
The meandering tails are consistent with instabilities expected for streaming
motions \citep[e.g.][]{Wareing07} or interaction at ionization front
\citep{Williams2003}, whilst solely shadowing by density enhancement 
requires straight tails by necessity.  (In some case ionization edges are visible within
the tails.)  Finally, the absence of clear tails in the outer regions is less
easily explained with shadowing. The issue is controversial, and further
velocity information would be required to resolve it.

As mentioned before, the knots change appreciably in appearance between the
inner and outer nebula. In the inner nebula, tails are always seen. In the
outer nebula, the knots appear more diffuse and tails are absent. This implies
a change in shaping mechanism. Whereas streaming motions appear important in
the inner region, the structures in the outer nebula suggest that it does not play
a role there.  Within the streaming model, the implication is that the knots
have a differential velocity with respect to the ambient medium in the inner
nebula, but not or less so in the outer nebula. A plausible model is that the
inner knots system is being overrun by faster expanding, lower density gas,
but that this has not yet reached the outer knot system.

 \citet{Meaburn98} find a difference in expansion velocity between the system
 of knots and the ionized gas, of 17\,km\,s$^{-1}$, with the knots moving
 slower.  The wind speed in the nebula shows significant variations, with
 velocities up to 50\,km\,s$^{-1}$ with respect to the nebula mean velocity
 \citep{Walsh87}.  The sound speed in the ionized nebula is of the order of
 10\,km\,s$^{-1}$. This is similar to the velocity differential: we may expect
 both sub-sonic and super-sonic winds, depending on the local conditions and
 on the flow velocity within the tail.

 If the gas flow models are correct, it points to the prevalent wind
  interaction being subsonic, because of the common presence of narrowing
  tails.  However, there are also cases of widening tails (knot f), which
  requires supersonic winds.  It would be of interest to measure detailed
  velocity fields for the tails and the surrounding gas. The tail velocities
  may possibly be measured from the H$_2$ lines.  

Some of the more complicated tail structures may be due to feedback, where an
ambient wind is disturbed by another knot.  The multiple brightness-peaks and
meandering tail in a knot (Fig.\,\ref{fig-indivisual}) may be explained
through the complex interaction between nearby cores \citep{Pittard05}.

Fig.\,\ref{fig-indivisual} shows that some knots have a dip in brightness in
the middle of the tail which remains unexplained. If this is due to the
pressure balance, the temperature would be lowest at this point and increase 
again further out (related to the shadowing by the knot). It could also be due
to a change to a supersonic flow at this point \citep{Pittard08}. Detailed
studies of temperature and velocity distribution along these tails may help
distinguish the various models, in so far impossible.

It does not appear possible to view the three models in isolation. In reality,
a combination of effects is likely required.
Streaming motions can help explain
the shapes, but the observed structures implies that streaming motions only
occur in the inner nebula, not the outer. The crescent shapes trace
photo-ionization, and photo-evaporation. Shadowing will occur, but its
importance to the shaping remains to be proven. Integrated models taken all
effects into account would be desirable.

\section{Conclusions}
We have obtained a wide field view image of the Helix nebula in excellent
seeing conditions.  Knots are found throughout the inner edge to the outer
ring, and H$_2$ is always associated with knots.  Chemical and PDR models do
not favour H$_2$ formation in the ionized and neutral gas in planetary
nebulae, suggesting that the observed H$_2$ gas is a remnant of molecular gas
formed during the AGB phase.  The MOIRCS image shows that knots come in a
variety of shapes.  These various knot morphologies suggest that the inner 
knots are being overrun, and shaped by, a faster wind. However, such a wind
would not yet have reached the outer system of knots,.

\acknowledgements
We appreciate technical support from the Subaru Telescope staff during the observations.
M.M. is a JSPS fellow.
M.M. appreciates discussions at UCL.
We also thank for Dr. J.M. Pittard for detailed explanations about his model.

{\it Facilities:} \facility{Subaru Telescope}, \facility{MOIRCS}

\clearpage






\end{document}